\documentclass[12pt,a4paper]{article}

\makeatletter
\@addtoreset{equation}{section}
\makeatother

\begin{document}
\begin{flushright}
\footnotesize
\footnotesize
CERN-TH/2000-096\\
{\tt hep-th/0003226}\\
March, $2000$
\normalsize
\end{flushright}

\begin{center}

\vspace{.8cm}
{\LARGE {\bf Non-BPS D-brane Solutions in Six}}\\
\vskip 7pt 
{\LARGE {\bf Dimensional Orbifolds}}

\vspace{1cm}


{\bf Yolanda Lozano}

\vspace{.1cm}

{
{\it Theory Division, CERN\\
1211 Gen\`eve 23, Switzerland}\\
{\tt yolanda.lozano@cern.ch}
}

\vspace{.4cm}

\vspace{2cm}


{\bf Abstract}

\end{center}
\begin{quotation}

\small

Starting with the non-BPS D0-brane solution of IIB/$(-1)^{F_L}I_4$
constructed recently by Eyras and Panda we construct via T-duality
the non-BPS D2-brane and D1-brane solutions of IIB/$(-1)^{F_L}I_4$
and IIA/$(-1)^{F_L}I_4$ predicted by Sen. The D2-brane couples
magnetically to the vector field of the NS5B-brane living in the
twisted sector of the Type IIB orbifold, 
whereas the D1-brane couples (electrically
and magnetically) to the self-dual 2-form potential of the NS5A-brane
that is present in the twisted sector of the Type IIA orbifold construction.
Finally we discuss the eleven dimensional interpretation of these branes
as originating from a non-BPS M1-brane solution of M-theory
orientifolded  by $\Omega_\rho I_5$.

\end{quotation}

\vspace{1cm}

\newpage

\pagestyle{plain}

\section{Introduction and Summary}

The study of brane-antibrane systems in string and M-theory has
brought a new perspective onto the understanding of 
non-perturbative extended objects (see 
\cite{Senrev}-\cite{BGrev}, and references therein).
BPS branes arise
as by-products when the tachyonic mode of the open
strings connecting the brane and the antibrane condenses in a 
vortex-like configuration. Also, one can deduce in this
framework the presence of new non-BPS extended objects 
when the tachyon condenses, instead, in a kink-like configuration.
These objects have subsequently been shown to play a key role 
in testing string dualities beyond the BPS level. 
In Type II and M-theory they are however unstable, because the open strings
ending on them contain real tachyonic excitations, but
the instability can be cured when the theory is projected out by
a certain symmetry that removes the tachyons from the 
spectrum.
This happens in particular for some non-BPS
branes in Type I
and in certain orbifold/orientifold constructions of Type
II and M-theory.

An interesting and still open problem is the construction of the 
supergravity solutions corresponding to these non-BPS branes.
In the Type II theories given that they are unstable
one does not expect to find stable classical solutions of the 
supergravity that could be associated to these branes.
Only in some cases it has been possible to construct classical solutions,
like the one corresponding to a Kaluza-Klein monopole-antimonopole
pair in M-theory, that is stabilized by suspending the system in an
external magnetic field (see \cite{Sen1} and \cite{Ro,Youm,CET,M,JM} 
for related work), or the solution of the, unstable, non-BPS D-instanton
of Type IIA constructed in \cite{HHK}. For the stable non-BPS branes
one however expects to find solutions of the supergravity equations
of motion that describe these branes in the strong coupling
regime.

Recently, using the boundary state formalism, Eyras and Panda \cite{EP}
found the asymptotic behavior of the solution corresponding to the
non-BPS D0-brane of the Type IIB theory orbifolded by
$(-1)^{F_L}I_4$ \cite{Sen2,BG}\footnote{Here $F_L$ denotes the left-moving
spacetime fermion number, and $I_4:x^i\rightarrow -x^i\,;i=1,\dots 4$.}.
This theory is S-dual to Type IIB orientifolded by
$\Omega I_4$, $\Omega$ being the worldsheet
parity reversal operation. The twisted sector of this orientifold
construction of Type IIB consists, in the compact case, on 16 O5 orientifold
fixed planes together with a D5-brane on top of each 
plane, each D5-O5 system carrying an SO(2)
vector potential. This theory contains
massive non-BPS states in the perturbative spectrum arising from
open strings stretched between a D5-brane and its image \cite{Sen*}. 
These states are stable, since they are the lightest states
charged under the SO(2) gauge
field of the twisted sector, and correspond in the strong coupling limit 
to the non-BPS D0-branes of IIB/$(-1)^{F_L}I_4$ \cite{Sen2,BG}.
These non-BPS D0-branes are also charged with respect to the
SO(2) vector field of the, S-dual, twisted sector, and this renders 
them stable.
Moreover, for a critical value of the radii of the compact orbifold
a pair of branes satisfies a no-force condition at least when the
distance is larger than the string scale 
\cite{GS,MOT}\footnote{In \cite{LS} it is shown that for coincident
branes a vacuum configuration in which the branes attract each other
seems to be more favourable even at the critical radius.
As mentioned in that reference
this result is however challenged by the fact that 
the expectation value for the tachyon fields is beyond the range 
of validity of the approximation. It would be very interesting
to clarify this point further.}. 
This opens the possibility of constructing solutions 
corresponding to a large number of parallel non-BPS D0-branes, 
which could correctly describe
the weak coupling regime of the theory. More importantly for our
discussion, it allows the possibility of constructing 
infinite arrays of non-BPS
D0-branes, from where one can derive other non-BPS D-brane
solutions via T-duality transformations.

In this paper we concentrate on this and other stable non-BPS
branes, that occur in the six dimensional orbifold/orientifold
constructions obtained by projecting out the Type IIB theory
by $\Omega I_4$ and duality-related operations. In particular,
we focus on the Type IIB and Type IIA theories divided out by
$(-1)^{F_L}I_4$. The $(-1)^{F_L}$ operation is identified in the
strong coupling limit as the transformation reversing the orientation
of open D-strings (D2-branes) in the Type IIB (Type IIA) theory,
as implied by its connexion via S-duality 
with the $\Omega$ symmetry of Type IIB \cite{Sen**}
(in Type IIA a further T-duality transformation is required).
Therefore, the twisted sector of Type IIB/$(-1)^{F_L}I_4$ 
can be described non-perturbatively, in the compact case, as
16 O5-NS5B systems \cite{Kuta,MS}, and that of Type IIA/$(-1)^{F_L}I_4$
as the same number of O5-NS5A systems \cite{Kuta}. 
An O5-plane with a NS5B-brane on top of it contains an SO(2)
gauge field associated to open D-strings stretched between the 
NS5-brane and its image,
and the non-BPS D0-branes are charged with respect to this twisted
field. In the Type IIA theory, the O5-NS5A system contains a
self-dual 2-form field associated to open D2-branes stretched between
a NS5A-brane and its image, 
and therefore the twisted sector contributes with 
SO(2) self-dual 2-form potentials\footnote{See \cite{Kuta} and
\cite{DM,W} for a more detailed description of the twisted sector.}.

In \cite{Senrev} Sen conjectured that together with the non-BPS D0-brane, 
coupled electrically
to the SO(2) vector field of the NS5B-O5 system, there is 
a non-BPS D2-brane, placed on the orbifold plane, that couples 
magnetically to the same vector field,
and arises from open D3-branes stretched between a NS5B-brane and
its mirror.
Also, T-duality predicts a non-BPS D1-brane in  
IIA/$(-1)^{F_L}I_4$ located on the orbifold plane, which should couple,
electrically and magnetically, to the self-dual 2-form potential of
the twisted sector, and arise from open D2-branes stretched between
a NS5A-brane and its mirror.

In this note we construct these D-brane solutions. We also show that
there is a non-BPS M1-brane solution of M-theory orientifolded by
$\Omega_\rho I_5$ \cite{DM,W},
where $\Omega_\rho$ reverses the orientation of 
the M2-brane, to which all these branes are related by reduction
and dualities. This provides a unifying picture within M-theory
of the stable non-BPS branes that occur in the six dimensional
orbifold/orientifold constructions related to $\Omega_\rho I_5$. 

M-theory on a 5-torus orientifolded by $\Omega_\rho I_5$ contains 
a twisted sector that can be identified 
as 16 O5-M5 systems, together with other 16 O5 orientifold fixed planes
which do not contribute any twisted states \cite{W}.
This theory contains non-BPS M1-branes that arise from open
M2-branes stretched between an M5-brane and its mirror \cite{Sen*}. 
They couple,
electrically and magnetically, to the self-dual 2-form potential
living in the M5-brane, what makes them stable.
In reducing to the Type IIA theory one can consider two
possibilities: 

\begin{enumerate}

\item Reduce along a worldvolume direction of the M5-O5 system. 
In this case one obtains Type IIA 
orientifolded by $\Omega I_5$, whose twisted sector states arise from
16 D4-O4 systems. This theory contains perturbative massive non-BPS 
states, which can be interpreted in M-theory as non-BPS 
M1-branes wrapped on the eleventh direction \cite{Sen*}, 
as well as non-perturbative
non-BPS strings coming from open D2-branes stretched between a D4-brane
and its mirror, which correspond to unwrapped M1-branes in M-theory
\cite{Senrev}. T-duality along one of
the orbifolded directions gives then
rise to the Type IIB theory orientifolded by $\Omega I_4$, whose twisted
sector is described by 16 D5-O5 systems. This theory contains 
perturbative non-BPS particle states and non-perturbative non-BPS 2-branes, 
connected by T-duality with the non-BPS objects of IIA.

\item Reduce M-theory/$\Omega_\rho I_5$ along
one of the orbifolded directions. In this case one obtains 
Type IIA projected out by $(-1)^{F_L}I_4$, with a twisted sector
consisting of 16 NS5-O5 systems.
The non-BPS M1-brane gives rise to a non-BPS D1-brane in IIA
that couples (electrically and magnetically) to the self-dual
2-form potential living in the worldvolume of the NS5A-O5. 
Now T-duality along a worldvolume direction of the NS5A-O5 maps
the theory onto Type IIB divided by $(-1)^{F_L}I_4$, with a twisted sector
identified as 16 NS5-O5 systems.  
Non-BPS D0-branes are coupled electrically to the SO(2) vector field of
the twisted sector, and non-BPS D2-branes
magnetically. These branes are related
to the non-BPS D1-brane by T-duality.

\end{enumerate}

\noindent Consistently with the whole duality picture \cite{HW}, 
the two theories that are obtained by either reducing along an
M5-brane direction and then T-dualizing along a transverse direction,
or viceversa, are related by S-duality. In particular one obtains
IIB/$\Omega I_4$ and IIB/$(-1)^{F_L}I_4$ respectively.
We also see that the non-BPS M1-brane of M-theory/$\Omega_\rho I_5$
is the eleven dimensional origin of the non-BPS branes that can be
defined in the
Type II orbifolds/orientifolds obtained by reduction.

\section{The non-BPS D0-brane solution of \cite{EP}}

The asymptotic behavior of the solution corresponding to the
non-BPS D0-brane of Type IIB orbifolded by $(-1)^{F_L}I_4$
has been derived in \cite{EP} using the boundary state formalism.
In this formalism one can compute the long distance behavior
of the massless fields generated by the D-brane and predict in
this manner the asymptotic form of the corresponding classical 
solution \cite{DFLPRS}. 
A pair of non-BPS D0-branes satisfies a no-force
condition when the orbifold is compactified to a particular 
critical value of the radii
\cite{GS,MOT}. When this happens it is possible to construct
periodic infinite arrays of non-BPS D0-branes
and compute T-dual solutions, which is what we shall be doing in the
next sections. Note that both the no-force condition and the validity
of the classical solution and the T-duality rules hold for distances
larger than the string scale.

The asymptotic form of the solution of \cite{EP}, corresponding to a
D0-brane situated at one of the fixed points of the orbifold,
reads, in string frame\footnote{In \cite{EP} a somewhat more general solution
depending on a free parameter $a$ is given, derived by
impossing the no-force condition of a pair of branes at the critical radii
as a constraint for the background fields. Here we have chosen to work
with the strictly linearized solution, though the same kind of 
generalization can be done for our solutions.}:

\begin{eqnarray}
\label{D0}
&&ds_{D0}^2= -(1-{1\over 3}
{\kappa_6 T_0 \over 2\pi^2\Omega_4}
{1 \over |y|^3}+\dots)dt^2+\nonumber\\
&&\nonumber\\
&&\hspace{1cm}+(1+{1\over 3}{\kappa_6 T_0 \over
2\pi^2\Omega_4}{1\over |y|^3}+\dots)\left(
\delta_{mn}dy^m dy^n+\right.
\nonumber\\
&& \nonumber\\
&&\hspace{1cm}\left.+\delta_{ij}dx^idx^j\right) \, ; \,\,m,n=1,\dots 5\, ;
\,\, i,j=1,\dots 4\, ,\nonumber\\
& &\nonumber\\
&&e^\phi = 1+{1\over 2}{\kappa_6 T_0\over 2\pi^2\Omega_4}
{1\over |y|^3}+\dots \, ,\nonumber\\
& &\nonumber\\
&&C^{(1)}_0=-{1\over 3} {\kappa_6 Q_0 \over \sqrt{2}\Omega_4}
{1\over |y|^3}+\dots\, .
\end{eqnarray}

\noindent Here we have taken $\alpha^\prime=1$ but otherwise the
notation is that in \cite{EP}. Namely, $\kappa^2_{D}=8\pi G_D$,
$\kappa^2_{D-d}=\kappa^2_{D}/V_d$, with $V_d$ the volume of the
$d$ dimensional space,
$\Omega_4$ is the area of a unit sphere surrounding the D0-brane, 
$T_0$ is the tension of the brane, $Q_0$ its 
charge\footnote{Impossing open-closed string consistency for 
boundary states, $T_0$ and $Q_0$ are fixed to: $T_0=8\pi^{7/2}$, 
$Q_0=8\sqrt{2}\pi^{3/2}$, (see \cite{EP}).}, 
$y^m$, $m=1,\dots 5$ the longitudinal directions along the NS5B-O5
worldvolume, and
$x^i$, $i=1,\dots 4$, the transverse, orbifolded directions. 
The critical value of the
radii: $R_c=1/\sqrt{2}$, has already been substituted in the solution.
$C^{(1)}$ is the vector potential coming from the twisted sector,
under which the D0-brane is charged.

\section{The non-BPS D1-brane of IIA/$(-1)^{F_L}I_4$}

Considering a periodic infinite array of non-BPS D0-branes
along the $y^5$ direction we can construct via T-duality 
a non-BPS D1-brane solution in the Type IIA theory
projected out by $(-1)^{F_L}I_4$. 
This brane is situated at one of the fixed points of the orbifold
with its worldsheet extended along the non-compact spacetime.
We find:

\begin{eqnarray}
\label{D1}
&&ds_{D1}^2=(1-\frac12\frac{\kappa_6 T_1}{2\pi^2 \Omega_3}
\frac{1}{|y|^2}+\dots)(-dt^2+d\sigma^2)+\nonumber\\
&&\nonumber\\
&&\hspace{1cm}+(1+\frac12\frac{\kappa_6 T_1}
{2\pi^2\Omega_3}\frac{1}{|y|^2}+\dots)\left(\delta_{mn}dy^m dy^n
+\delta_{ij}dx^i dx^j\right) \, ; \nonumber\\
&&\nonumber\\
&&\hspace{1cm}m,n=1,\dots 4\, ; \,\, i,j=1,\dots 4\, ,\nonumber\\
& &\nonumber\\
&& e^\phi = 1+\frac12\frac{\kappa_6 T_1}{2\pi^2\Omega_3}\frac{1}{|y|^2}
+\dots \, ,\nonumber\\
& &\nonumber\\
&& C^{(2)}_{0\sigma}=\frac12 \frac{\kappa_6 Q_1}{\sqrt{2}\Omega_3}
\frac{1}{|y|^2}+\dots \, .
\end{eqnarray}

\noindent
Here $T_1$ is the tension of the brane, $Q_1$ its charge and
$\Omega_3$ the area of the unit 3-sphere surrounding the string.
The twisted sector consists on a 2-form potential, under which the
D1-brane is charged. T-duality implies that this field
is to be interpreted as the 2-form potential living in a NS5A-O5
system.

This D1-brane solution can be interpreted as a ten dimensional 1-brane
located at the origin of the four-dimensional compact space.
Considering first a single compactified direction, a 1-brane sitting
at the origin of the $S^1$ can be seen from the point of view of the
covering space of the $S^1$ 
as an equally spaced array of D1-branes in the $S^1$ direction. If
$\vec{x}$ denotes a vector in the full eight dimensional transverse
space, we can then approximate\footnote{The choice of this function will 
be clear below.}
$1/|x|^6$ by a sum $\sum_{n\in Z}1/(r^2+(x^4-2\pi n
R_4)^2)^3$, with $r^2=\sum_{m=1}^4 y^m y_m+\sum_{i=1}^3 x^i x_i$, 
and, assuming that the size of the compact direction is smaller than
the distance in the non-compact space, we can further approximate the
sum by an integral. Repeating this process for the four compact directions
we can finally write (see 
for instance \cite{Ricc} for more details):

\begin{equation}
\frac{1}{|y|^2}\approx \frac{1}{|x|^6}\Pi_{i=1}^4 (2\pi R_i)
(I_1 I_2 I_3 I_4)^{-1}\, ,
\end{equation}

\noindent where $R_i$ are the radii of the compactified orbifold and
$I_n\equiv\int_0^{\pi}d\theta \sin^n{\theta}$.
Substituting back in the expression for the D1-brane solution in the
compact orbifold we can then obtain the corresponding 
solution in the uncompactified
case:

\begin{eqnarray}
\label{D1unc}
&&ds_{D1}^2=(1-\frac16 \frac{\kappa_{10}T_1}{\Omega_7}\frac{1}{|x|^6}
+\dots)(-dt^2+d\sigma^2)+\nonumber\\
&&\nonumber\\
&&\hspace{1cm}+(1+\frac16 \frac{\kappa_{10}T_1}{\Omega_7}
\frac{1}{|x|^6}+\dots)\left(\delta_{mn}dy^m dy^n 
+\delta_{ij}dx^i dx^j\right) \, ;\nonumber\\
&&\nonumber\\
&&\hspace{1cm}m,n=1,\dots 4\, ;\,\, i,j=1,\dots 4\, ,\nonumber\\
& &\nonumber\\
&& e^\phi=1+\frac16 \frac{\kappa_{10}T_1}{\Omega_7}\frac{1}{|x|^6}
+\dots \, ,\nonumber\\
& &\nonumber\\
&& C^{(2)}_{0\sigma}=\frac12 \frac{\kappa_6 Q_1}{\sqrt{2}\Omega_3}
\frac{1}{|y|^2}+\dots \, .
\end{eqnarray}

\noindent Now $\Omega_7$ is the
area of the unit 7-sphere surrounding the string.
The expression for the 2-form potential is the same as in the compactified
case since it lives in the twisted sector.
In \cite{EP} it is shown that for the non-BPS D0-brane this kind of 
approach relating the uncompactified and the compactified solutions
gives the same answer than the boundary state analysis  
of the uncompactified orbifold. This should be the case also for the
D1-brane. 

It is straightforward to check that this D1-brane solution 
solves the equations of motion derived from
an action $S_{{\rm untwisted}}+S_{{\rm twisted}}$, where:

\begin{equation}
S_{{\rm untwisted}}=\frac{1}{2\kappa^2_{10}}\int d^{10}x\,
e^{-2\phi}\sqrt{|{\rm det}g|}\left( R+4 (\partial\phi)^2\right)\, ,
\end{equation}

\noindent and the action corresponding to the twisted sector is 
proportional to the worldvolume effective action associated to 
a pair NS5A-O5. This
reads, in string frame and to first order in $\alpha^\prime$:

\begin{equation}
\begin{array}{rcl}
S_{{\rm twisted}}&\sim & \int d^6 y \sqrt{|{\rm det}g|}
\left(1+({\cal H}^{(3)})^2+\dots\right) - 
\int d^6 y \sqrt{|{\rm det}g|}= \\
&&\\
&=&\int d^6 y \sqrt{|{\rm det}g|} ({\cal H}^{(3)})^2+\dots \, .
\end{array}
\end{equation}

\noindent Here ${\cal H}^{(3)}$ is the field strength of the
self-dual 2-form potential of the NS5A-brane, and the self-duality 
condition is impossed at the level of the 
equations of motion. The
metric is restricted to the position of the orientifold fixed plane.

\section{The non-BPS D2-brane of IIB/$(-1)^{F_L}I_4$}

Performing now a T-duality transformation of the D1-brane solution
along the $y^4$ direction
we obtain a D2-brane solution of IIB/$(-1)^{F_L}I_4$. 
This brane is located at one of the fixed points of the orbifold,
with its worldvolume extending along the non-compact spacetime.
Taking
the non-BPS D1-brane in the compactified orbifold at the
critical radii, where it is possible to
construct periodic infinite arrays of strings,
and applying the T-duality rules we find:

\begin{eqnarray}
\label{D2}
&&ds_{D2}^2=(1-\frac{\kappa_6T_2}{2\pi^2\Omega_2}\frac{1}{|y|}
+\dots)(-dt^2+d\sigma_1^2+d\sigma_2^2)+\nonumber\\
&&\nonumber\\
&&\hspace{1cm}+(1+\frac{\kappa_6T_2}
{2\pi^2\Omega_2}\frac{1}{|y|}+\dots)\left(\delta_{mn}dy^m dy^n 
+\delta_{ij}dx^i dx^j\right) \, ;\nonumber\\
&&\nonumber\\
&&\hspace{1cm}m,n=1,\dots 3\, ;\,\, i,j=1,\dots 4\, ,\nonumber\\
& &\nonumber\\
&& e^\phi=1+\frac12\frac{\kappa_6 T_2}{2\pi^2\Omega_2}\frac{1}{|y|}
+\dots \, ,\nonumber\\
& &\nonumber\\
&& C^{(3)}_{0\sigma_1\sigma_2}=-\frac{\kappa_6 Q_2}
{\sqrt{2}\Omega_2}
\frac{1}{|y|}+\dots \, .
\end{eqnarray}

\noindent Here $T_2$ is the tension of the brane, $Q_2$ its charge and
$\Omega_2$ the area of the unit 2-sphere surrounding the 2-brane. 

The same analysis of the previous section gives the following form
for the solution in the uncompactified case:

\begin{eqnarray}
\label{D2unc}
&&ds_{D2}^2=(1-\frac15 \frac{\kappa_{10}T_2}{\Omega_6}\frac{1}{|x|^5}
+\dots)(-dt^2+d\sigma_1^2+d\sigma_2^2)+\nonumber\\
&&\nonumber\\
&&\hspace{1cm}+(1+\frac15 \frac{\kappa_{10}T_2}
{\Omega_6}\frac{1}{|x|^5}+\dots)\left(\delta_{mn}dy^m dy^n
+\delta_{ij}dx^i dx^j\right) \, ;\nonumber\\
&&\nonumber\\
&&\hspace{1cm}m,n=1,\dots 3\, ;\,\,i,j=1,\dots 4\, ,\nonumber\\
& &\nonumber\\
&& e^\phi=1+\frac{1}{10} \frac{\kappa_{10}T_2}{\Omega_6}\frac{1}{|x|^5}
+\dots \, ,\nonumber\\
& &\nonumber\\
&& C^{(3)}_{0\sigma_1\sigma_2}=-\frac{\kappa_6 Q_2}
{\sqrt{2}\Omega_2}
\frac{1}{|y|}+\dots \, .
\end{eqnarray}

\noindent Now $\Omega_6$ is the area of the unit 6-sphere surrounding
the membrane.

This brane is electrically charged with respect to the 3-form potential
of the NS5B-O5 system, or equivalently, magnetically charged with
respect to its vector potential. Therefore, it solves the equations of
motion derived from $S_{{\rm untwisted}}+S_{{\rm twisted}}$, with:

\begin{equation}
\begin{array}{rcl}
S_{{\rm twisted}}&\sim & \int d^6 y\sqrt{|{\rm det}g|}
\left(1+({\tilde {\cal F}}^{(4)})^2+\dots\right)-\int d^6 y
\sqrt{|{\rm det}g|}=\\
&&\\
&=&\int d^6 y\sqrt{|{\rm det}g|}({\tilde {\cal F}}^{(4)})^2+\dots\, ,\\
\end{array}
\end{equation}

\noindent where we have dualized the vector field of the NS5B-brane
onto a 3-form potential with field strength
${\tilde {\cal F}}^{(4)}$, and the metric is restricted to the position
of the orientifold fixed plane.

\section{The non-BPS M1-brane of M-theory/$\Omega_\rho I_5$}

Oxidizing the D1-brane solution of the Type IIA theory on the orbifold
we can obtain the expression for a stable M1-brane solution of
M-theory orientifolded by $\Omega_\rho I_5$. In the compact case we
obtain:

\begin{eqnarray}
\label{M1com}
&&d{\hat s}_{M1}^2=(1-\frac56 \frac{\kappa_6 T_1}{2\pi^2\Omega_3}
\frac{1}{|y|^2}+\dots)(-dt^2+d\sigma^2)+\nonumber\\
&&\nonumber\\
&&\hspace{1cm}+(1+\frac16 \frac{\kappa_6 T_1}{2\pi^2\Omega_3}
\frac{1}{|y|^2}+\dots)\left(\delta_{mn}dy^mdy^n+\delta_{ij}dx^i
dx^j\right)+\nonumber\\
&&\nonumber\\
&&\hspace{1cm}+(1+\frac23 \frac{\kappa_6 T_1}{2\pi^2\Omega_3}
\frac{1}{|y|^2}+\dots)dz^2\, ;\nonumber\\
&&\nonumber\\
&&\hspace{1cm}m,n=1,\dots 4\, ;\,\,i,j=1,\dots 4\, ,\nonumber\\
&&\nonumber\\
&& {\hat C}^{(2)}_{0\sigma}=\frac12 \frac{\kappa_6 Q_1}{\sqrt{2}\Omega_3}
\frac{1}{|y|^2}+\dots\, .
\end{eqnarray}

\noindent This solution has an $SO(1,1)\times SO(4)\times SO(4)
\times U(1)$ symmetry, i.e. it corresponds to an asymmetric orbifold.
This is in agreement with the results
of \cite{DM,W}, which show that the orbifold corresponding to
M-theory on $T^5/\Omega_\rho I_5$
has to be asymmetric 
so that the twisted sector cancels the gravitational anomalies
involved in the construction. We see that asymptotically, 
i.e. in the region where the dilaton of the Type IIA theory is of
order 1, the orbifold regains isotropy, also in agreement with
\cite{DM,W}.  

Uplifting the D1-brane solution corresponding to Type IIA on the
uncompactified orbifold we find an M1-brane on $R^{1,5}\times 
(R^4\times S^1)/\Omega_\rho I_5$: 

\begin{eqnarray}
\label{M1}
&&d{\hat s}_{M1}^2=(1-\frac{5}{18} \frac{\kappa_{10}T_1}{\Omega_7}
\frac{1}{|x|^6}
+\dots)(-dt^2+d\sigma^2)+\nonumber\\
&&\nonumber\\
&&\hspace{1cm}+(1+\frac{1}{18} \frac{\kappa_{10}T_1}
{\Omega_7}\frac{1}{|x|^6}+\dots)\left(\delta_{mn}dy^m dy^n+\delta_{ij}
dx^idx^j\right)
 +\nonumber\\
&&\nonumber\\
&&\hspace{1cm}+(1+\frac29 \frac{\kappa_{10}T_1}{\Omega_7}
\frac{1}{|x|^6}+\dots)dz^2\, ;\nonumber\\
&&\nonumber\\
&&\hspace{1cm}m,n=1,\dots 4\, ;\,\,i,j=1,\dots 4\, ,\nonumber\\
& &\nonumber\\
&& {\hat C}^{(2)}_{0\sigma}=\frac12 \frac{\kappa_6 Q_1}{\sqrt{2}\Omega_3}
\frac{1}{|y|^2}+\dots \, .
\end{eqnarray}

\noindent As we did in the previous sections we can 
interpret this solution as an
M1-brane located at the origin of the $z$-circle, and find the
expression for the corresponding solution in the completely
uncompactified case:

\begin{eqnarray}
\label{M1unc}
&&d{\hat s}_{M1}^2=(1-\frac{5}{21} \frac{{\hat \kappa}_{11}{\hat T}_1}
{\Omega_8}\frac{1}{|x|^7}+\dots)(-dt^2+d\sigma^2)+\nonumber\\
&&\nonumber\\
&&\hspace{1cm}+(1+\frac{1}{21} \frac{{\hat \kappa}_{11}{\hat T}_1}
{\Omega_8}\frac{1}{|x|^7}+\dots)\left(\delta_{mn}dy^m dy^n+\delta_{ij}
dx^idx^j\right)+\\
&& \nonumber\\
&&\hspace{1cm}+(1+\frac{4}{21} \frac{{\hat \kappa}_{11}{\hat T}_1}
{\Omega_8}
\frac{1}{|x|^7}+\dots)dz^2\, .\nonumber
\end{eqnarray} 
 
\noindent Here $\Omega_8$ is the area of the unit 8-sphere 
surrounding the string, and we have used 
${\hat \kappa}_{11}=\kappa_{10}\,(2\pi R_z)^{1/2}$,
${\hat T}_1=T_1 \,{\hat \kappa}_{11}/\kappa_{10}$. 
${\hat C}^{(2)}$ remains the same since it lives
in the twisted sector.

Finally, the contribution to the supergravity action from the twisted sector
is proportional to the worldvolume action describing an M5-O5 system, 
which in quadratic approximation reads:

\begin{equation}
{\hat S}_{{\rm twisted}}\sim\int d^6 {\hat y} \sqrt{|{\rm det}
{\hat g}|} ({\hat {\cal H}}^{(3)})^2+\dots
\end{equation}

\noindent Here ${\hat {\cal H}}^{(3)}$ is the field strength associated
to the self-dual 2-form potential of the M5-brane worldvolume, with
the self-duality condition impossed at the level
of the equations of motion, and the metric is restricted to the position
of the orientifold fixed plane.

\subsection*{Acknowledgements}

It is a pleasure to thank Laurent Houart for very interesting discussions,
and Eduardo Eyras for pointing out some mistakes in a previous version
of this paper.

\end{document}